\documentclass[12pt,a4paper]{article}

\usepackage{amssymb,amsfonts,amsmath}
\usepackage{xcolor}

\title{Conditional probability framework for entanglement and its decoupling from tensor product structure}

\author{Irina Basieva and Andrei Khrennikov\\
International Center for Mathematical Modeling\\
 in Physics and Cognitive Sciences \\
 Linnaeus University,  V\"axj\"o, Sweden}

\date{}

\begin{document}

\maketitle

\abstract{Our aim is to make a step towards clarification of foundations for the notion of entanglement (both physical and mathematical) by representing it in the conditional probability framework. In Schr\"odinger's words, this is entanglement of knowledge which can be extracted via conditional measurements. In particular, quantum probabilities are interpreted as conditional ones (as, e.g., by Ballentine). We restrict considerations to perfect conditional correlations (PCC) induced by measurements 
(``EPR entanglement''). Such entanglement is coupled to the pairs of observables  with the projection type state update as the back action of measurement. In this way, we determine a special class of entangled states. One of our aims is to decouple the notion of entanglement from the compound systems. The rigid association of entanglement with the state of a few body systems stimulated its linking with quantum nonlocality (``spooky action at a distance''). However, already by Schr\"odinger entanglement was presented as knotting of knowledge (about statistics) for one observable $A$ with knowledge about another observable $B.$}    
    
\section{Introduction}

In recent articles \cite{NL1}--\cite{HP}, the nonlocal interpretation of the violation of the Bell inequalities \cite{Bell0}--\cite{CHSH} was criticized; it was shown that, in fact, this violation can be simply obtained by taking into account incompatibility (or mathematically non-commutativity \rm{NC})  of observables involved into the Bell tests. We point out that the quantum nonlocality interpretation of violation of the Bell type inequalities was criticized by many authors.\footnote{Articles \cite{NL1, NL2, HP} are close to the papers expressing (in one or another way) the  viewpoint that the seed of violation of the Bell inequalities is in incompatibility, not in nonlocality. I would especially highlight the papers which explored quantum formalism  \cite{Loub}--\cite{PGF}. Other authors explored incompatibility indirectly (i.e., without coupling to quantum formalism) as nonexistence of the joint probability distribution  \cite{Cetto0}--\cite{Boughn1a}.} However, the idea about quantum nonlocality is not solely coupled to the violation of  the Bell inequalities. Befittingly, it is present in the formalism of quantum mechanics and encoded in the projection postulate when this postulate is applied to the entangled states of compound systems. This feature of the quantum state update (resulting from the back action of measurement) was firstly discussed in the EPR paper \cite{EPR}. But its authors only briefly mentioned nonlocality as an alternative to incompleteness of quantum mechanics. An interesting discussion connecting the EPR projection argument with the Bell inequality argument can be found in Aspect's papers \cite{AA0,AA1}. He pointed out that in principle one need  the Bell inequality tests to only to put the projection type nonlocality into the experimental framework. Thus, irrespectively to the Bell inequality, the notion of entanglement in combination with the projection postulate has the flavor of nonlocality. 
\footnote{Note that the difference between the notions of Bell locality, EPR locality and nonsignaling was first specified
mathematically in article \cite{LE1}. In the present paper we discuss EPR locality. See also \cite{LE2}-- \cite{LE3} for Bell locality and nonlocality.}

In this paper we clarify of the notion of entanglement. The key point is the presentation of  entanglement  in the {\it  conditional probability framework.} In Schr\"odinger's words \cite{SCHE, SCHE1}, this is entanglement of knowledge (entanglement of predictions) which can be extracted via conditional measurements. In particular, quantum probabilities are interpreted as conditional probabilities (cf. Koopman \cite{Koopman}, Ballentine \cite{BL, BL1}, see also \cite{INT, KHRB2, KHRB3, INT1}). We restrict considerations to {\it perfect conditional correlations} (PCC) induced by measurements ({\it ``EPR-entanglement''} \cite{EPR}). Such entanglement is coupled to the pairs of observables  with the projection type state update as the back action of measurement \cite{VN, Luders}. In this way, we determine a special class of entangled states. 

One of our aims is to decouple the notion of entanglement from the compound systems. The rigid association of entanglement with the state of a few body systems (originated in Schr\"odinger's article \cite{SCHE, SCHE1}) stimulated its linking with quantum nonlocality. However, already in \cite{SCHE, SCHE1}  entanglement was presented as knotting of knowledge (about statistics) for one observable $A$ with knowledge about another observable $B.$  This coupling gives the possibility to gain knowledge about $B$ through $A$-measurement
(see also Bohr \cite{BR}). An entangled state $|\psi \rangle$ served as a part of the mathematical formalism for such conditional extraction of knowledge. Our framework is decoupled from the tensor product structure of the state space related to compound systems.
The tensor product expression of  entanglement is considered as a special mathematical representation.  Even CHSH inequality can be analyzed without consideration of compound systems (see \cite{NL1,NL2}).
  
In the presented mathematical framework, it is meaningless to speak about entanglement without pointing to the concrete pairs of observables $A_i, B_i, i=1,...,n,$  for which PCC yields (cf. Bohr \cite{BR}).  In this aspect our paper is close to the approach presented in the papers \cite{Z1,Z2}; we cite \cite{Z2}: 	

{\it ``Here we propose that a partitioning of a given Hilbert space is induced by the experimentally accessible 
observables .... In this sense entanglement is always relative to a particular set of experimental capabilities.''}

This viewpoint on entanglement differs crucially from the standard definition of entanglement; see, e.g., \cite{WER}:

{\it ``A state is said to be entangled if it cannot be written as a convex sum of tensor product states.''} 

The observable based viewpoint on entanglement matches better the views of Bohr who treated QM as measurement theory
\cite{BR0}; in fact, Bohr's reply to Einstein \cite{BR} was in the line with the observational interpretation of entanglement, but the reasoning of Einstein and his coauthors was firmly based on the tensor-product interpretation. May be this striking difference in 
views on entanglement was one of the reasons for their misunderstanding.   

We shall discuss similarities and dissimilarities between the frameworks of the present paper and articles \cite{Z1,Z2} 
in appendix A.
\footnote{The main difference is that our paper is directed to foundational clarification of the notion of entanglement through exploration of the calculus of conditional probabilities (a la Ballentine \cite{BL, BL1}). Thus, conditional probability and probability update are the key structures in our study. There is nothing about them in the aforementioned articles. They are directed merely to algebraic aspects of entanglement structures.} 

EPR-entanglement is represented with the simple systems of linear equations for  projections. PCC-states can be determined not only for compatible observables $A, B,$ but even for incompatible ones. 
Then the order of conditioning is important, say PCCs $A=+, B=+$  and $B=+, A=+$   are not equivalent. Of course, the formalism is essentially simplified for compatible observables. Therefore the main body of the paper is devoted to the latter case, i.e., 
$A, B,$ are mathematically represented by Hermitian operators $\hat A, \hat B$ and $[\hat A, \hat B]=0.$  And usual consideration of 
dichotomous observables $A, B= \pm 1,$ i.e., $\hat A^2=I, \hat B^2=I$ makes PCC-theory even simpler and clearer.       

In parallel to mathematical reformulation of the notion of entanglement, we clarify the meaning of the projection postulate as the state representation of probability update generated by the back action of measurement (see article \cite{NL2} for details, see also Ballentine \cite{BL, BL1}) - the quantum analog of the classical Bayesian probability update. 

The paper also contains a brief foundational discussion (section \ref{BFD}). In short, the main foundational implication of the conditional probability approach to PCC is disillusion of the EPR notion of an element of reality.
    
\section{The projection postulate as the mathematical tool for quantum probability conditioning}

In the quantum formalism an observable $A$ (with a discrete range of values) is represented by a Hermitian operator 
\begin{equation}
\label{L2a}
\hat A= \sum_\alpha \alpha E^A(\alpha),
\end{equation} 
where $E^A(\alpha)$ is projection onto the space ${\cal H}_A(\alpha)$ of eigenvectors for the eigenvalue $\alpha.$ For a pure state $\vert \psi \rangle,$ the probability to get the outcome $A=\alpha$ is given by the Born's rule: 
\begin{equation}
\label{L2a1}
p(A=\alpha | \psi)= \Vert  E^A(\alpha)\vert \psi \rangle \Vert^2. 
\end{equation} 
A measurement with the  outcome $A= x$ generates back-action onto system's state:
\begin{equation}
\label{L2b}
\vert \psi \rangle \to \vert \psi \rangle_\alpha^A=  E^A(\alpha)\vert \psi \rangle/ \Vert E^A(\alpha) \vert \psi \rangle \Vert.
\end{equation} 
This is the projection postulate in the L\"uders form \cite{Luders}. \footnote{It is often called the von Neumann projection postulate. However, von Neumann used the projection postulate only to observables represented by operators  with non-degenerate spectra; for operators  with degenerate spectra von Neumann considered more general state transformations; in particular, updating of a pure state given by  a vector $\vert \psi \rangle$ can lead to a mixed state given by a density operator $\hat \rho$ \cite{VN}. Later von Neumann considerations were formalized within theory of quantum instruments describing state updates of non-projection type even for observables with non-degenerate spectra (see, e.g.,  \cite{DV}--\cite{Ariano}). In \cite{VNL}, it was shown that appealing to non-projection transformations has interesting foundational insights on the basic problems of quantum foundations.} 

The projection postulate is the mathematical tool for quantum conditioning. Measurements of another observable, say $B,$ conditioned on 
the outcome $A=\alpha$ lead to probability (see, e.g., \cite{KHR_CONT}): 
\begin{equation}
\label{Lx3m}
P(B=  \beta| A= \alpha, \psi)= \Vert  E_B(\beta)  \vert \psi_A^{\alpha} \rangle\Vert^2=\frac{ \Vert  E_B(\beta)   E_A(\alpha) \vert \psi \rangle \Vert^2}{\Vert E_A(\alpha) \vert \psi \rangle\Vert^2}.
\end{equation}
We shall use this definition of conditional probability to define the special form of entanglement, 
corresponding to PCC.  

\section{EPR-entanglement}

We would like to embed the perfect EPR correlations in the framework of quantum conditioning which is mathematically described with the projection postulate - probability one conditioning. The framework is more general than the one needed for the EPR situation and includes even conditioning for pairs of incompatible observables $A$ and $B,$ i.e., $[\hat A, \hat B]\not=0.$ However, the framework of conditional EPR entanglement has a rich structure only in the case of compatible observables, i.e., $[\hat A, \hat B] =0.$ Morover, our construction is not reduced to the state spaces with the tensor product structure. The latter is considered as just an illustrative example. Thus, as was pointed out in introduction, we eliminate rigid coupling of entanglement with tensor product and compound systems. 

\subsection{Perfect conditional correlations}

Consider two discrete  observables $A$ and $B$ which are represented by Hermitian operators $\hat{A}$ and  $\hat{B}$ with the spectral decompositions:
\begin{equation}
\label{Lx1}
\hat{A}= \sum_\alpha  \alpha E_A(\alpha), \; 
\hat{B}=\sum_\beta \beta E_B(\beta).
\end{equation}
It is also assumed that, for these observables,  the state updates as the back action of measurement is mathematically described with the projection postulate
\begin{equation}
\label{Lx2}
\vert \psi \rangle  \to \vert \psi_A^{\alpha} \rangle =    E_A(\alpha) \vert \psi \rangle/ \Vert E_A(\alpha) \vert \psi \rangle\Vert,
\end{equation}
\begin{equation}
\label{Lx2az}
\vert \psi \rangle  \to \vert \psi_B^{\beta} \rangle =    E_B(\beta) \vert \psi \rangle/ \Vert E_B(\beta) \vert \psi \rangle\Vert .
\end{equation}
We remark that such updates are possible not an arbitrary state, but only if  
\begin{equation}
\label{Lx2a}
E_A(\alpha) \vert \psi \rangle \not=0, \mbox{i.e.}, \; P(A= \alpha|\psi) \not=0, 
\end{equation}
\begin{equation}
\label{Lx2b}
E_B(\beta) \vert \psi \rangle \not= 0, \mbox{i.e.}, \; P(B= \beta|\psi) \not=0 
\end{equation}
We are interested in a state $\vert \psi \rangle$  satisfying the following condition.  The conditional probability to get the outcome  $B=\beta$ if  the preceding $A$-measurement had the outcome $A=\alpha$ equals to 1,
\begin{equation}
\label{Lx3}
P(B=  \beta| A= \alpha, \psi)= \frac{ \Vert  E_B(\beta)   E_A(\alpha) \vert \psi \rangle \Vert^2}{\Vert E_A(\alpha) \vert \psi \rangle\Vert^2} =  1.
\end{equation}

\medskip

{\bf Definition 1.}  {\it If equality (\ref{Lx3}) holds, then, in state $\vert \psi \rangle,$  the observables are perfectly  conditionally correlated (PCC) for the values  $(A=\alpha, B=\beta).$}\footnote{Here the order is important: first $A-$measurement and then $B$-measurement. As we shall see, it is important even in the case of compatible observables.}  

\medskip

Hence, we get the equation for the state $\vert \psi \rangle$ 
\begin{equation}
\label{Lx4}
\Vert  E_B(\beta)   E_A(\alpha) \vert \psi \rangle \Vert^2  = \Vert E_A(\alpha) \vert \psi \rangle\Vert^2 .
\end{equation}
We remark that 
\begin{equation}
\label{Lx4bb}
E_A(\alpha) \vert \psi \rangle= \sum_y E_B(y) E_A(\alpha) \vert \psi \rangle,
\end{equation}
hence
\begin{equation}
\label{Lx5}
 \Vert E_A(\alpha) \vert \psi \rangle \Vert^2 = \sum_y \Vert E_B(y) E_A(\alpha) \vert \psi \rangle\Vert^2
= \Vert  E_B(\beta)   E_A(\alpha) \vert \psi \rangle \Vert^2.
\end{equation}
Hence, $\Vert E_B(y) E_A(\alpha) \vert \psi \rangle\Vert^2=0$ for any $ y \not= \beta.$ And equality (\ref{Lx4bb}) is reduced 
to the equality: 
\begin{equation}
\label{Lx6}
E_B(y) E_A(\alpha) \vert \psi \rangle = 0 \; \mbox{for \; all} \; y \not= \beta, 
\end{equation}
or 
\begin{equation}
\label{Lx7}
E_B(\beta) E_A(\alpha) \vert \psi \rangle = E_A(\alpha) \vert \psi, 
\end{equation}
i.e., $E_A(\alpha) \vert \psi\rangle$ is the eigenvector of $E_B(\beta)$ with eigenvalue $\lambda=1.$ 
The conditions (\ref{Lx2a}), (\ref{Lx7}) give  the description of the PCC-states for $(A=\alpha, B= \beta).$

We start to study properties of PCC-states by considering unitary transformations, so let $\hat U : {\cal H} \to  {\cal H}$ be a unitary transformation. Set $\vert \psi \rangle_U=  \hat U \vert \psi \rangle$ and $\hat A_U= \hat U A \hat U^\star, 
\hat B_U= \hat U B \hat U^\star.$ Then $E_{B_U}(\beta) E_{A_U}(\alpha) \vert \psi \rangle_U=  \hat U E_B(\beta) E_A(\alpha) \vert \psi \rangle = \hat U E_A(\alpha)\hat U^\star   \hat U \vert \psi \rangle = E_{A_U}(\alpha) \vert \psi \rangle_U.$ 
We also remark that $E_{A_U}(\alpha) \vert \psi \rangle_U =  \hat U E_A(\alpha) \vert \psi \rangle \not=0.$

Hence, unitary transformation applied consistently to the state and operators preserves the PCC-property.  

\medskip

We now reformulate this condition in terms of eigensubspaces, set ${\cal H}_A(\alpha) = E_A(\alpha){\cal H}, 
{\cal H}_B(\beta) = E_B(\beta) {\cal H}.$ Then (\ref{Lx7}) can be rewritten as  
\begin{equation}
\label{Lx7g}
E_A(\alpha) \vert \psi \rangle  \in {\cal H}_B(\beta).
\end{equation}
We remark that if 
\begin{equation}
\label{Lx7gg}
{\cal H}_A(\alpha) \subset {\cal H}_B(\beta),
\end{equation}
then this condition is satisfied for any state $\vert \psi\rangle$ such that (\ref{Lx2a}) holds. 

\subsection{Common $(A=\alpha, B=\beta)$ eigenvectors }

Equality (\ref{Lx7}) implies that the vector $\vert \phi \rangle = E_A(\alpha) \vert \psi\rangle$ is the common eigenvector
of projections $E_A(\alpha)$ and $E_B(\beta)$ (in particular, 
of operators $\hat A$ and $\hat B),$ since  $E_A(\alpha) \vert \phi \rangle = \vert \phi \rangle$ per definition  and 
$E_B(\beta) \vert \phi \rangle = \vert \phi \rangle$ per (\ref{Lx7}). Thus, in particular, $[\hat A, \hat B] \vert \phi\rangle =0.$
This fact devalues consideration of the above scheme in the noncommutative case.  

We illustrate the previous statement by the following example. 
Consider the four dimensional case. Let $\{e_1,e_2,e_3, e_4\}$ be some orthonormal basis in  ${\cal H}.$ 
Let ${\cal H}_A(\pm)$  be generated by $\{e_1,e_2\}, \{e_3, e_4\}.$ Set 
$f_1=(e_1+ e_3)/\sqrt{2}, f_2=(e_2+ e_4)/\sqrt{2}, f_3 =(e_1 - e_3)/\sqrt{2}, , f_4=(e_2- e_4)/\sqrt{2};$
let  ${\cal H}_B(\pm)$  be generated by $\{f_1,f_2\}, \{f_3, f_4\}.$ Then say $E_B(+)$ and $E_A(+)$ do not have a common
eigenvector, therefore  $(A=+, B=+)$ PCC-states do not exist.

Now let us proceed another way around. Let state $\vert \phi \rangle$ belong to subspace ${\cal H}_{AB}(\alpha, \beta)= {\cal H}_A(\alpha) \cap {\cal H}_B(\beta)$ which is nontrivial subspace of ${\cal H}.$ Then, this is $(\alpha, \beta) $PCC-state for the observables $A,B.$ 

Do all PCC-states belong to subspace ${\cal H}_{AB}(\alpha, \beta)?$

 The answer is not. Take nonzero vectors $|\phi \rangle \in  {\cal H}_{AB}(\alpha, \beta)$ and  $\vert \xi\rangle$ which is orthogonal to subspace ${\cal H}_A(\alpha).$ Set $\vert \psi\rangle= |\phi \rangle +   \vert \xi\rangle.$ Then $E_A(\alpha) \vert \psi\rangle= |\phi \rangle =  E_B(\beta) E_A(\alpha) \vert \psi\rangle.$

\subsection{Entanglement w.r.t. a set of values}

Consider now another pair, $(\alpha^\prime, \beta^\prime),$ of the values of the observables $A,B.$ The PCC-condition for  
$\alpha^\prime, \beta^\prime$  has the form: 
\begin{equation}
\label{Lx7rr}
E_B(\beta^\prime) E_A(\alpha^\prime) \vert \psi \rangle = E_A(\alpha^\prime) \vert \psi \rangle. 
\end{equation}
We also have additional condition (\ref{Lx2a}) for the value $\alpha^\prime.$
There can be found PCC-states   for  both  pairs 
$(\alpha, \beta)$ and $(\alpha^\prime, \beta^\prime).$ 

More generally consider observables with values $(\alpha_i)$ and $(\beta_i)$ and some set $\Gamma$ of pairs $(\alpha_i, \beta_j).$   

\medskip

{\bf Definition 2.}  (EPR entanglement) {\it If, for all pairs from the set $\Gamma,$   $\vert \psi \rangle$ is PCC-state, then such state is called EPR $\Gamma$-entangled.}

\medskip

So, conditions (\ref{Lx3}) and  (\ref{Lx7}) hold for all pairs $(\alpha, \beta) \in \Gamma.$

We would be mainly interested in sets $\Gamma$ such that each of $\alpha$ and $\beta$ values appears in pairs once and only once
(see section \ref{DDD}  for discussion). We call such {\it EPR entanglement complete.}

For example, for two dichotomous observables with $\alpha, \beta=\pm 1,$ 
we consider, e.g.,  the set of the  pairs $(A=+, B=-), (A=-, B=+),$ in short, EPR $(\pm, \mp)$ entanglement, 
or the  pairs $(A=+, B=+), (A=-, B=-),$ EPR $(\pm, \pm)$ entanglement; see section \ref{LL} for  
dichotomous compatible observables. (We repeat that generally the order of measurements is important.)
      
\section{Order of measurements, symmetric EPR entanglement}

For incompatible observables (represented by noncommuting operators, $[\hat{A},\hat{B}] \not=0),$ the   order of measurements
is important.

Suppose now that the observables are compatible, i.e.,  $[\hat{A},\hat{B}] =0.$ One might expect that the order of measurements is unimportant for the perfect correlation. The situation is more complicated.   We proceed under the probability non-degeneration conditions (\ref{Lx2a}), 
(\ref{Lx2b}). Then PCCs $(A=\alpha, B= \beta)$ and $(B=\beta, A=\alpha)$ are expressed by the pair of equalities:
\begin{equation}
\label{ya}
E_B(\beta)   E_A(\alpha) \vert \psi \rangle = E_A(\alpha) \vert \psi \rangle,
\end{equation}
\begin{equation}
\label{yb}
E_A(\alpha)E_B(\beta)  \vert \psi \rangle = E_B(\beta) \vert \psi \rangle.
\end{equation}
They imply that 
\begin{equation}
\label{Lx3z2}
E_A(\alpha) \vert \psi \rangle =  E_B(\beta) \vert \psi \rangle .
\end{equation}

The latter equality implies that $E_B(\beta) E_A(\alpha) \vert \psi \rangle = E_B(\beta) E_B(\beta) \vert \psi \rangle=
E_B(\beta) \vert \psi \rangle,$ i.e., equality (\ref{ya}); in the same way one obtains   (\ref{yb}). Thus, in the commutative case
the condition of order irrelevance of PCC is given by the equality (\ref{Lx3z2})  (and the equalities (\ref{Lx2a}),  (\ref{Lx2b})).  

For such state $\vert \psi \rangle,$ the $A=\alpha$ outcome implies  the $B=\beta$ outcome with probability one and vice verse, i.e., for  $\vert \psi \rangle,$  PCC is symmetric w.r.t. order of measurements.
We call such EPR entanglement symmetric.\footnote{ Here the ordering is not due to incompatibility, but  due to temporal structure of experiment. Which observable is measured first?   We remark that this is the standard situation in  the Bell inequalities tests. Here the probability of time coincidence for  $A$- and $B$-outcomes is zero. Approximate coincidence is determined via using the time window, but the real experimental data shows which click was the first and which was the second.} 

\medskip

{\bf Definition 3.} (Symmetric PCC and entanglement) {\it If $\vert \psi \rangle$ is a PCC-state for $(A=\alpha_j, B=\beta_i)$ and  $(B=\beta_i, A=\alpha_j)$ for all pairs from set $\Gamma,$ then such  state is called symmetrically EPR $\Gamma$-entangled.}

\section{Dichotomous compatible observables}
\label{LL}

Consider now compatible dichotomous observables $A, B= \pm 1.$ We study, e.g., $(A= +, B= -), (A= -, B=+)$ EPR   
entanglement. This is complete entanglement involving all possible values of observables. 
We proceed under conditions (\ref{Lx2a}), (\ref{Lx2b}) guarantying the existence of conditional probabilities 
for $\alpha, \beta=\pm 1.$

First we show that this is symmetric EPR entanglement, it does not depend on order.  We show that  
PCC for $(A= +, B= -)$ implies PCC for $(B= +, A= -)$ as well as PCC for $(A= -, B= +)$ implies PCC for $(B=-, A= +)$ and vice verse.

Thus $(A= +, B= -), (A= -, B=+)$ and  $(B= -, A= +), (B=+, A= -)$  EPR entanglements are equivalent,  
so we can simply speak about $A=-B$ entanglement. 

PCC for $(A= +.B= -)$  is characterized by the equality  
\begin{equation}
\label{Lx9}
E_B(-) E_A(+) \vert \psi \rangle = E_A(+) \vert \psi\rangle. 
\end{equation}
Hence, 
$(I-  E_B(+)) (I-  E_A(-)) \vert \psi \rangle = (I- E_A(-)) \vert \psi\rangle,$ or 
$E_B(+) E_A(-) \vert \psi \rangle = E_B(+)  \vert \psi \rangle,$ and now we use commutativity:
\begin{equation}
\label{Lx9x}
E_A(-) E_B(+)  \vert \psi \rangle = E_B(+)  \vert \psi \rangle.
\end{equation}
This is the PCC-condition for $(B=+, A= -).$

In the same way, PCC for  $(A=  -, B= +)$ is characterized by the equality
\begin{equation}
\label{Lx9a}
E_B(+) E_A(-) \vert \psi \rangle = E_A(-) \vert \psi \rangle. 
\end{equation}
implies the equality 
\begin{equation}
\label{Lx9xa}
E_A(+) E_B(-)  \vert \psi \rangle = E_B(-)  \vert \psi \rangle,
\end{equation}
the latter is the PCC-condition for $(B= - , A=+ ).$ 

The equalities (\ref{Lx9}), (\ref{Lx9xa}) imply the order irrelevance equality
\begin{equation}
\label{Lx9t}
E_A(+) \vert \psi\rangle = E_B(-)  \vert \psi \rangle, 
\end{equation}
which in turn implies these inequalities. In the same way, the equalities (\ref{Lx9x}), (\ref{Lx9a}) are equivalent to 
the equality
\begin{equation}
\label{Lt}
E_A(-) \vert \psi\rangle = E_B(+)  \vert \psi \rangle, 
\end{equation}

We also remark that the equality (\ref{Lx9t}) is equivalent to the equality (\ref{Lt}). Hence, each of them characterizes 
both $(A= \pm+, B= \mp)$ and $(B=\pm, A= \mp)$ EPR entanglements, so we can simply speak about 
$A=-B$ EPR entanglement characterized by the equivalent conditions
\begin{equation}
\label{Lx10}
E_B(+) \vert \psi \rangle=   E_A(-) \vert \psi \rangle (\not=0),  \; E_B(-) \vert \psi \rangle = E_A(+) \vert \psi \rangle  (\not=0).
\end{equation}

We remark that $A=-B$ EPR entanglement implies equality of the probabilities :
\begin{equation}
\label{Lx3f}
P(A=+ 1 |\psi) =  P(B= - 1|\psi) (\not=0), \; P(A= -1 |\psi) =P(B= + 1|\psi) (\not=0).
\end{equation}

We also have 
$\hat A\hat B= E_A(+) E_B(+)+ E_A(-) E_B(-) - E_A(+) E_B(-)- E_A(-) E_B(+),$ and 
$E_A(+) E_B(+)|\psi\rangle = E_A(+) E_A(-)|\psi\rangle=0,  E_A(-) E_B(-)|\psi\rangle = E_A(-) E_A(+)|\psi\rangle=0,$
and $(E_A(+) E_B(-) \psi\rangle + E_A(-) E_B(+) \psi\rangle= (E^2_A(+)  + E_A^2(-))\psi\rangle = \psi\rangle,$ i.e. 
\begin{equation}
\label{nu1}
\hat A\hat B |\psi\rangle = - |\psi\rangle,
\end{equation}
i.e. $|\psi\rangle$ is an eigenvector the product of operators
with the  eigennvalue $\lambda= -1.$ Hence correlation of observables in this state equals to -1, 
\begin{equation}
\label{nu2}
\langle A B\rangle_\psi \equiv \langle \psi| \hat A\hat B| \psi\rangle = -1. 
\end{equation}
In the same way for the EPR $A=B$ entanglement, 
\begin{equation}
\label{nu3}
\hat A\hat B |\psi\rangle =  |\psi\rangle,
\end{equation}
and 
\begin{equation}
\label{nu4}
\langle A B\rangle_\psi \equiv \langle \psi| \hat A\hat B| \psi\rangle = +1. 
\end{equation}

\subsection{Examples}
\label{EX}

\subsubsection{$\rm{dim}\; {\cal H}= 5$} 
Consider  the five dimensional case. 
Let $\{e_1,e_2,e_3, e_4, e_5\}$ be some orthonormal basis in  ${\cal H}.$ 
Let ${\cal H}_A(\pm)$  and ${\cal H}_B(\pm)$ be generated by $\{e_1,e_2,e_3\}, \{e_4, e_5\}$ and $\{e_1,e_2\}, \{e_3, e_4,e_5\},$ 
respectively. An arbitrary $\vert \psi\rangle$ can be expended as 
\begin{equation}
\label{nt}
\vert \psi\rangle = \sum_i c_i e_i, \;   \sum_i |c_i|^2 =1.
\end{equation}
Let us find EPR entangled states  for $A=-B.$  
Then $E_A(-) \vert \psi \rangle= c_4 e_4 + c_5 e_5$ and $E_B(+) \vert \psi \rangle= c_1 e_1 + c_2 e_2.$ 
Thus, the only state which satisfy equation $E_A(-) \vert \psi \rangle= E_B(+) \vert \psi \rangle$ is the state 
$\vert \psi \rangle= e_3.$ However, $E_A(-) \vert \psi \rangle=0.$  There are no EPR entangled states for $A=-B.$ 

Let us find EPR entangled states  for $A=B.$ We have $E_A(+) \vert \psi \rangle= c_1 e_1 + c_2 e_2 +c_3 e_3$ and $E_B(+) \vert \psi \rangle= c_1 e_1 + c_2 e_2.$ Hence, $c_3=0,$ the EPR 
entangled state has the form
\begin{equation}
\label{nt1}
\vert \psi\rangle = (c_1 e_1 + c_2 e_2) + (c_4 e_4 + c_5 e_5), 
\end{equation}
where $|c_1|^2 + |c_2|^2 \not=0, |c_4|^2 + |c_5|^2 \not=0.$ 
We remark that vectors $|\phi_{++}\rangle = c_1 e_1 + c_2 e_2$ and $|\phi_{--}\rangle = c_4 e_4 + c_5 e_5$ 
belong, respectively,  to the subspaces of joint eigevectors ${\cal H}_{AB}(++)$ and  ${\cal H}_{AB}(--)$
(see section \ref{ZZZ}).
 
\subsubsection{$\rm{dim}\; {\cal H}= 4$}
\label{H4}

Consider $A=-B$ EPR entanglement in the four dimensional case. 
Let $\{e_1,e_2,e_3, e_4\}$ be some orthonormal basis in  ${\cal H}.$ 
Let ${\cal H}_A(\pm)$  and ${\cal H}_B(\pm)$ be generated by $\{e_1,e_2\}, \{e_3, e_4\}$ and $\{e_1,e_3\}, \{e_2, e_4\},$ 
respectively. An arbitrary $\vert \psi\rangle$ can be expended as (\ref{nt}).

Then $E_A(-) \vert \psi \rangle= c_3 e_3 + c_4 e_4$ and $E_B(+) \vert \psi \rangle= c_1 e_1 + c_3 e_3.$   These equalities with 
(\ref{Lx10}) imply that  $c_1=0, c_4=0,$ i.e., $\vert \psi\rangle = c_2 e_2 + c_3 e_3, \; |c_2|^2 +  |c_3|^2 = 1.$ Now encode the basis vectors by using 
the values of observables, i.e., $e_1=\vert ++\rangle, e_2= \vert +-\rangle, e_3= \vert - +\rangle, e_4= \vert --\rangle.$ Hence,
\begin{equation}
\label{e1}
\vert \psi\rangle = c_{+-} \vert +-\rangle + c_{-+}  \vert -+\rangle, \; \; c_{+-}, c_{-+} \not=0, 
\end{equation}
(the last inequalities are due to conditions (\ref{Lx2a}), (\ref{Lx2b})); here $|c_{+-}|^2 + |c_{-+}|^2=1.$

In the same way,  $A=B$ EPR entanglement restricts the class of states to    
\begin{equation}
\label{e2}
\vert \psi\rangle = c_{++} \vert ++\rangle + c_{--}  \vert --\rangle,  \; \; c_{++}, c_{--} \not=0. 
\end{equation}

Consider now the case  ${\cal H}_A(\pm)$  and ${\cal H}_B(\pm)$ are generated by $\{e_1,e_2,e_3\}, \{e_4\}$ and $\{e_1\}, \{e_2,e_3, e_4\},$  respectively. 

Consider $A=-B$ EPR entanglement. Then $E_A(-) \vert \psi \rangle= c_4 e_4$ and $E_B(+) \vert \psi \rangle= c_1 e_1.$ Hence, $c_1=0, c_4=0,$ and $E_A(-) \vert \psi \rangle= E_B(+) \vert \psi =0,$ i.e., the corresponding conditional probabilities are no well defined. In this case entangled states do not exist. Similarly,  let ${\cal H}_A(\pm)$  and ${\cal H}_B(\pm)$ be generated by $\{e_1,e_2,e_3\}, \{e_4\}$ and $\{e_1, e_2, e_3\}, \{e_4\},$  respectively. Then $E_A(-) \vert \psi \rangle= c_4 e_4$ and $E_B(+) \vert \psi \rangle= c_1 e_1+ c_2 e_2+c_3 e_3.$ In this case entangled states do not exist.  We note that in this space 
encoding of eigenvectors with values of the operators is not one to one, $e_1=\vert ++\rangle, e_2= \vert +-\rangle, e_3= \vert +-\rangle, e_4= \vert --\rangle.$ 

Consider now $A=B$ EPR entanglement; $E_A(-) \vert \psi \rangle= c_4 e_4$ and $E_B(-) \vert \psi \rangle= c_2 e_2 + c_3 e_3 + c_4 e_4.$
Hence, $c_2=0, c_3=0.$ And  $\vert \psi \rangle= c_1 e_1 + c_4 e_4,$ where $c_1 \not=0, c_4 \not=0.$

\subsubsection{\rm{dim}\; ${\cal H}= 3$}

Consider  $A=-B$ entanglement in the case $\rm[{dim} \; = 3.$ Let  ${\cal H}_A(\pm)$  and ${\cal H}_B(\pm)$ are generated 
by $\{e_1,e_2\}, \{e_3\}$ and $\{e_1\}, \{e_2, e_3\}.$ We remark that these vectors can be encoded as
$e_1=\vert ++\rangle, e_2= \vert +-\rangle, e_3= \vert - +\rangle$ (although  ${\cal H}$ is not a tensor product space).\footnote{If we consider the above example with $\rm{dim}\; {\cal H}= 5,$ then such encoding is impossible, since it would lead to coding:
 $e_1=\vert ++\rangle, e_2= \vert + +\rangle, e_3= \vert +-\rangle, e_4= \vert - -\rangle, e_5= \vert --\rangle.$}

We have $E_A(-) \vert \psi \rangle= c_3 e_3$ and $E_B(+) \vert \psi \rangle= c_1 e_1.$ Thus, 
$\vert \psi \rangle = e_2= \vert  +- \rangle.$ But, this vector does not satisfy the probability non-degeneration condition, since
$E_A(-) \vert +- \rangle =0.$ This, vector generates the perfect conditional correlation $A=+1, B= -1,$ but not $A=-1, B= + 1.$ Thus there are no $(\pm \mp)$  entangled states.    
%$E_A(+) \vert \psi \rangle= c_1 e_1 + c_2 e_2$ and $E_B(-) \vert \psi \rangle= c_2 e_2 + c_3 e_3.$ 

Now, we study $A=B$ entanglement.  let  ${\cal H}_A(\pm)$  and ${\cal H}_B(\pm)$ are generated by $\{e_1,e_2\}, \{e_3\}$ and 
$\{e_1\}, \{e_2, e_3\}.$ Then $E_A(-) \vert \psi \rangle= c_3 e_3$ and $E_B(-) \vert \psi \rangle= c_2 e_2 + c_3 e_3.$ Hence, the entangled state  has the form 
$$ \vert \psi \rangle= c_1 e_1 + c_3 e_3= c_{++} \vert ++ \rangle + c_{--} \vert -- \rangle.$$ As always, $c_{++}, c_{--} \not=0.$ 

\subsubsection{\rm{dim}\; ${\cal H}= 2$} 

Consider  $A=-B$ entanglement in the case $\rm[{dim} \; {\cal H}= 2.$ Let  ${\cal H}_A(\pm)$  and ${\cal H}_B(\mp)$ are generated by $\{e_1\}, \{e_2\},$ respectively. Then, for $\vert \psi \rangle = c_1 e_1 +c_2 e_2, \; E_A(+)= c_1 e_1, E_B(-) \vert \psi \rangle=    c_1 e_1.$ Thus any state with $c_j \not=0$ is  $A=-B$ entangled.  

\subsubsection{The case of $\hat A = \hat B$}

Let us consider the case of the trivial entanglement, i.e., let $\hat A = \hat B.$ Since 
${\cal H}= {\cal H}_A(-) \oplus {\cal H}_A(+),$ then each state can be represented as
$\vert \psi \rangle=\vert \psi \rangle_- +  \vert \psi \rangle_+$ with the orthogonal components 
$\vert \psi \rangle_{\pm} \in {\cal H}_A(\pm).$ For $A=A$ EPR entanglement, the only restriction is the condition 
guaranteeing the existence of conditional probabilities, i.e.,    $\vert \psi \rangle_{\pm} \not=0.$ 

And it is clear that there are no $A=- A$ entangled states.  

\subsection{Tensor product state space} 
\label{TPS}

Consider the state space of a compound system $S=(S_1,S_2)$ given by the tensor product
 ${\cal H}={\cal H}_1 \otimes {\cal H}_2.$ Consider two dichotomous observables $a$ and $b$ for subsystems 
$S_1$ and $S_2$ given by Hermitian operators $\hat a = E_a(+) - E_a(-)$ and $\hat b = E_b(+) - E_b(-);$
set $\hat A = \hat a \otimes I= E_A(+) - E_A(-), E_A(\pm)= E_a(\pm) \otimes I$ and 
$\hat B =  I \otimes \hat b= E_B(+) - E_B(-), E_B(\pm)= I \otimes E_b(\pm).$ Set ${\cal H}_{1 \pm}= E_a(\pm) {\cal H}_{1},$ 
i.e., ${\cal H}_{1} = {\cal H}_{1-} \oplus {\cal H}_{1 +}$ and ${\cal H}_{2 \pm}= E_b(\pm) {\cal H}_{2},$ 
i.e., ${\cal H}_{2} = {\cal H}_{2-} \oplus {\cal H}_{2 +}.$  Also set ${\cal H}_{AB}(\alpha \beta)= 
E_A(\alpha) E_B(\beta) {\cal H}.$

Let $(|\alpha j\rangle)$ and  $(|\beta i\rangle)$ be orthonormal bases in spaces ${\cal H}_{1\alpha}$ and ${\cal H}_{2 \beta},$
composed of eigenvectors of the operators $\hat a$ and $\hat b,$ i.e.,  $ \hat a |\alpha j\rangle = \alpha |\alpha j\rangle, j=1,...,n_\alpha,$ and  $\hat b |\beta i\rangle= \beta |\beta i\rangle, i=1,...,m_\beta,$ where $\alpha, \beta= \pm.$ 
Then $(|\alpha j\rangle \otimes |\beta i\rangle \equiv |\alpha \beta ji \rangle),$ is the orthonormal basis in the space ${\cal H}.$ An arbitrary state in  ${\cal H}$ has the form:
$$
\vert \psi \rangle= \sum_{\alpha \beta ji} c_{\alpha \beta ji} |\alpha \beta ji \rangle . 
$$
Let us find the  EPR  entangled states for $A=-B,$  
$$
E_A(-) \vert \psi \rangle = \sum_{- \beta ji} c_{- \beta ji} |- \beta ji \rangle=  E_B(+) \vert \psi \rangle=
\sum_{\alpha + ji} c_{\alpha + ji} | \alpha + ji \rangle.
$$
Hence, all  coefficients $c_{- - ji}, c_{+ + ji}$ equal to zero, and 
\begin{equation}
\label{Ltp}
\vert \psi \rangle= \sum_{- + ji} c_{- + ji} |- + ji \rangle +  \sum_{+ - ji} c_{+ - ji} |+ - ji \rangle.
\end{equation}
We also have the non-degeneration conditions  
$$
\Vert E_A(-) \vert \psi \rangle \Vert^2=   \sum_{- + ji} \vert c_{- + ji} \vert^2 \not= 0, 
\Vert E_B(+) \vert \psi \rangle \Vert^2 = \sum_{+- ji} \vert c_{+ - ji} \vert^2 \not= 0,
$$
In superposition (\ref{Ltp}),   $c_{- + ji}\not=0$ and  $c_{+ - km}\not=0$ for at least one pair of indexes. We also note that 
the first sum belongs to the space ${\cal H}_{AB}(-+)$ and the second one to the space ${\cal H}_{AB}(+-),$ the joint eigenspaces 
of the operators (see section \ref{ZZZ}). 
  
Similarly we can describe all  $A=B$ EPR entangled states
\begin{equation}
\label{Ltp1}
\vert \psi \rangle= \sum_{++ ji} c_{+ + ji} |+ + ji \rangle +  \sum_{- - ji} c_{- - ji} |- - ji \rangle,
\end{equation}
where $c_{+ + ji}\not=0$ and  $c_{- - km}\not=0$ for at least one pair of indexes. 

Consider two dimensional Hilbert space $H$ and the Hermitian operators
$\hat a, \hat b$  in it with eigen-bases $(\vert f_{\pm}\rangle), \hat a \vert f_{\pm}\rangle = 
\pm \vert f_{\pm}\rangle,$ and $(\vert g_{\pm}\rangle), \hat b \vert g_{\pm}\rangle = \pm \vert g_{\pm}\rangle.$
Consider ${\cal H}= H \otimes H$ and operators $\hat A= \hat a \otimes I$ and $\hat B=I \otimes \hat b.$ 
Consider $(\pm, \mp)$ EPR entangled states, they have the form 
\begin{equation}
\label{e3}
\vert \psi \rangle= c_{+-} \vert f_+ g_- \rangle + c_{-+} \vert f_- g_+ \rangle, \; c_{+ -},  c_{- +}\not=0.  
\end{equation}
For example, let $x=(x_1, x_2,x_3), y=(y_1,y_2, y_3)$ and $\hat a= x_1 \sigma_1 + x_2 \sigma_2 + x_3 \sigma_3,
\hat b= y_1 \sigma_1 + y_2 \sigma_2 + y_3 \sigma_3$ be two Pauli vectors, then by selecting the correspodning 
bases of eigenvectors for them, one can write the EPR entangled state for them.   

Now the previous considerations were about the operators $A$ and $B$ respecting the tensor product structure of 
state space ${\cal H}.$ The EPR entangled states are entangled in the usual sense (but not vice verse).\footnote{Turn to the four dimensional case. Consider e.g. the state  
$$
\vert \psi\rangle =  c_{++} \vert ++\rangle + c_{+-} \vert +-\rangle + c_{--}  \vert --\rangle,
$$
with nonzero coefficients. Then it is not factorisable, i.e., entangled, but it is not EPR entangled.   
} However, 
in ${\cal H}$ one can consider operators which do not have the form 
$\hat A = \hat a \otimes I, \hat B = I \otimes \hat b.$ Then the corresponding EPR entangled states need not be 
entangled in the ordinary sense. 

\subsection{Joint eigenvectors and entanglement}
\label{ZZZ}

Consider again the general case of compatible dichotomous observables $A, B= \pm 1,$ i.e., without exploring the tensor product structure.  We study, e.g., $A=-B$ EPR   
entanglement. The state space can be represented as the direct sum of subspaces for the joint eigenvectors, $\hat A \vert \phi_{\alpha \beta}\rangle=|\phi_{\alpha \beta}\rangle, \hat B \vert \phi_{\alpha \beta}\rangle= |\phi_{\alpha \beta}\rangle,\alpha, \beta = \pm,$   
\begin{equation}
\label{Lm1}
{\cal H} = {\cal H}_{AB}(++) \oplus {\cal H}_{AB}(+-)\oplus {\cal H}_{AB}(-+)  \oplus {\cal H}_{AB}(--).
\end{equation}  

For any (normalized) vector 
\begin{equation}
\label{Lm27}
\vert \psi\rangle = \vert \psi_{+-}\rangle  + \vert \psi_{-+}\rangle,
\end{equation}
 where  
\begin{equation}
\label{Lm28}
\vert \psi_{\pm \mp}\rangle \in {\cal H}_{AB}(\pm \mp) \; \mbox{and} \; \vert \psi_{\pm \mp}\rangle \not=0,
\end{equation}  
 we have  $E_B(+) \vert \psi\rangle =  \vert \psi_{-+}\rangle= E_A(-) \vert \psi\rangle$ and 
$E_B(-) \vert \psi\rangle =  \vert \psi_{+-}\rangle= E_A(+) \vert \psi\rangle.$ Thus superposition of (non-zero) vectors from subspaces
${\cal H}_{AB}(+-), {\cal H}_{AB}(-+)$ is $A=-B$ EPR entangled.

Moreover, any $A=-B$ EPR entangled state is given by such a superposition. To prove this, suppose that 
\begin{equation}
\label{Lm2}
\vert \psi\rangle =  \vert \psi_{+-}\rangle  + \vert \psi_{-+}\rangle + 
\vert \psi_{++}\rangle  + \vert \psi_{--}\rangle,
\end{equation}
with components belonging to the corresponding subspaces.  Then, e.g.,  
$$
E_B(-) \vert \psi\rangle= \vert \psi_{+-}\rangle + \vert \psi_{--}\rangle, 
E_A(+) \vert \psi\rangle= \vert \psi_{+-}\rangle + \vert \psi_{++}\rangle.
$$
Hence, $\vert \psi_{--}\rangle=0, \vert \psi_{++}\rangle=0.$

Thus, for commuting dichotomous observables, the $A=-B$ EPR entangled  states are superpositions of joint eigenvectors corresponding to the pairs of eigenvalues $(\pm, \mp);$ see also examples in sections \ref{EX}, \ref{TPS}. The direct sum decomposition (\ref{Lm1}), in particular, implies that if some joint eigenspace is trivial, then the corresponding EPR entangled states do not exist. Say, if  ${\cal H}_{AB}(++)=\{0\},$ then there are no $A=B$ states.\footnote{We also note that if 
${\cal H} = {\cal H}_{AB}(++) \oplus {\cal H}_{AB}(--),$ then simply $\hat A= \hat B;$ if 
${\cal H} =  {\cal H}_{AB}(+-)\oplus {\cal H}_{AB}(-+),$ then $\hat A= - \hat B.$}

We remark that in our theory, the tensor product case is only a particular case. Generally {\it the EPR entanglement is characterized not by the tensor product, but by the direct sum operation.} The condition of non-factorization w.r.t. the tensor product is changed to the condition of non-triviality of state's components belonging to the corresponding joint eigensubspaces; say for $A=-B$ EPR entanglement, is characterized by conditions (\ref{Lm27}), (\ref{Lm28})

\subsection{Embedding into tensor product}

The consideration of section \ref{ZZZ} suggests the following embedding of the state space into tensor product.

To illustrate this construction, let us consider  the five dimensional case (section \ref{EX}). 
Let $\{e_1,e_2,e_3, e_4, e_5\}$ be some orthonormal basis in  ${\cal H}.$ 
Let ${\cal H}_A(\pm)$  and ${\cal H}_B(\pm)$ be generated by $\{e_1,e_2,e_3\}, \{e_4, e_5\}$ 
and $\{e_1,e_2\}, \{e_3, e_4,e_5\},$  respectively. We encode them as  
\begin{equation}
\label{Lma1}
e_1=\vert ++ \rangle_1, e_2=\vert ++\rangle_2, e_3=\vert +- \rangle, e_4=\vert -- \rangle_1, e_5=\vert -- \rangle_2.
\end{equation}

Here joint egeinspaces are generated by the following bases, 
\begin{equation}
\label{Lmb1}
{\cal H}_{AB}(++): (\vert ++ \rangle_1, \vert ++\rangle_2); 
{\cal H}_{AB}(+-): \vert +- \rangle; 
{\cal H}_{AB}(--): (\vert -- \rangle_1, \vert -- \rangle_2)
\end{equation} 
and ${\cal H}_{AB}(-+)=\{0\}.$  Hence, 
$$
{\cal H} = {\cal H}_{AB}(++) \oplus {\cal H}_{AB}(+-)\oplus  {\cal H}_{AB}(--).
$$
In particular, this automatically implies that there are no $A=-B$ entangled states.

Consider now two dimensional spaces with bases labeled as follows
$$
{\cal H}_a(+): \vert + \rangle_{1a}, \vert + \rangle_{2a},  
{\cal H}_a(-): \vert - \rangle_{1a}, \vert -\rangle_{2a} 
$$
and one dimensional spaces
$$
{\cal H}_b(+): \vert + \rangle_b, {\cal H}_b(-):  \vert -\rangle_b 
$$
Set ${\cal H}_a= {\cal H}_a(+) \oplus {\cal H}_a(-), {\cal H}_b= {\cal H}_b(+) \oplus {\cal H}_b(-),$ and 
${\bf H}= {\cal H}_a \otimes {\cal H}_b.$ The latter has the basis labeled as
$$
\vert ++\rangle_1 = \vert +\rangle_{1a} \vert + \rangle_b , 
\vert ++\rangle_2= \vert +\rangle_{2a} \vert + \rangle_b, 
\vert +- \rangle_1 =\vert +\rangle_{1a} \vert - \rangle_b,
\vert +- \rangle_2= \vert + \rangle_{2a} \vert - \rangle_b ,  
$$
$$
\vert - + \rangle_1= \vert - \rangle_{1a} \vert +\rangle_b, 
\vert - + \rangle_2 = \vert - \rangle_{2a} \vert + \rangle_b, 
\vert -- \rangle_1 = \vert - \rangle_{1a} \vert - \rangle_b, 
\vert --\rangle_2 = \vert - \rangle_{2a} \vert - 1\rangle_b.
$$
Tnen we can embed ${\cal H}$ into ${\bf H}$ in two ways. The vectors $e_1, e_2$ and $e_4, e_5$ are mapped in the corresponding 
vectors of ${\bf H}.$ But the vector $e_3$ can be mapped either to $\vert +- \rangle_1$ or to $\vert +- \rangle_2.$\footnote{ 
In the same way, we can proceed with construction based on two dimensional spaces ${\cal H}_b(\pm)$ and one dimensional 
spaces ${\cal H}_a(\pm).$} 
The operators $\hat A$ and $\hat B$ can be extended onto  ${\bf H}$
\begin{equation}
\label{e37}
\hat  {\bf A} |\alpha \beta\rangle_i= \alpha |\alpha \beta \rangle_i, \; 
 \hat {\bf B}  |\alpha \beta \rangle_i = \beta  |\alpha \beta\rangle_i.
\end{equation}
Hence, ${\cal H}_{{\bf A}}(\pm)$ are generated by vectors 
$$(\vert ++\rangle_1,
\vert ++\rangle_2, 
\vert +- \rangle_1,
\vert +- \rangle_2), \; (\vert -+\rangle_1,
\vert -+\rangle_2, 
\vert -- \rangle_1,
\vert -- \rangle_2)$$ and ${\cal H}_{{\bf A}}(\pm)$  by vectors 
$$(\vert ++\rangle_1,
\vert ++\rangle_2, 
\vert -+ \rangle_1,
\vert - +\rangle_2) , \;  
(\vert +-\rangle_1,
\vert +-\rangle_2, 
\vert -- \rangle_1,
\vert -- \rangle_2).
$$
Let us find EPR entangled states for ${\bf A} = - {\bf B};$ we have
$$
E_{\hat  {\bf A}} (+) \vert \psi\rangle= c_{++1} \vert ++\rangle_1 +
c_{++2} \vert ++\rangle_2 +  c_{+-1} \vert +- \rangle_1  + c_{+-2} \vert +- \rangle_2
$$
and 
$$
E_{\hat  {\bf B}} (-) \vert \psi\rangle= c_{+-1} \vert +-\rangle_1 +
c_{+-2} \vert +-\rangle_2 +  c_{--1} \vert -- \rangle_1  + c_{--2} \vert -- \rangle_2.
$$
Hence coefficients $c_{++i}, c_{--i}, i=1,2,$ are equal to zero and the EPR entangled state 
has the form
$$
\vert \psi\rangle=   c_{+-1} \vert +- \rangle_1  + c_{+-2} \vert +- \rangle_2 +  
c_{-+1} \vert - + \rangle_1  + c_{-+2} \vert - +\rangle_2,
$$  
where $ |c_{+-1}|^2 + |c_{+-2}|^2 \not=0, \; |c_{-+1}|^2 + |c_{-+2}|^2 \not=0$ 
and vector's norm equals to one.

Hence, as could be expected, tensor product embedding crucially changes the EPR entangled features. This shows restriction 
of the tensor product construction. 

\section{Can  the same outcome be perfectly conditionally correlated with two different outcomes?}
\label{DDD}

We again consider the general case, i.e., the operators do not need to be commuting. 

\subsection{$(A= \alpha, B=\beta)$ and $(A= \alpha, B=\beta^\prime).$}
\label{ABB}

Can a state satisfy the conditions of perfect  conditional correlations for the pairs 
$(\alpha, \beta)$ and  $(\alpha, \beta^\prime),$ where $\beta^\prime\not= \beta?$ The answer is no. Really, let
$$
E_B(\beta) E_A(\alpha) \vert \psi \rangle = E_A(\alpha) \vert \psi \rangle, \; 
E_B(\beta^\prime) E_A(\alpha) \vert \psi \rangle = E_A(\alpha) \vert \psi\rangle.
$$
Hence, $E_A(\alpha) \vert \psi\rangle =0.$ This contradicts to the necessary condition for definition of conditional 
probabilities.  So, it cannot happen  that by getting one fixed outcome $A=\alpha,$ one can get with probability  1 two different outputs of $B.$  For dichotomous variables, $\alpha, \beta = \pm 1,$ EPR entanglement of the form $(A=+, B=+$ and  
$(A=+, B=-)$ or $(A=-, B=+)$ and $(A=+, B=-).$ 

\subsection{$(A= \alpha, B=\beta)$ and $(A= \alpha^\prime, B=\beta).$}

Now turn to the question on the possibility to have both $(\alpha, \beta)$ and $(\alpha^\prime, \beta)$ perfect conditional  correlations for  $\alpha\not= \alpha^\prime.$
For $\gamma= \alpha, \alpha^\prime,$ we have
\begin{equation}
\label{Lx8n}
E_B(\beta) E_A(\gamma) \vert \psi \rangle = E_A(\gamma) \vert \psi \rangle, \; 
\end{equation}
In general, these two conditions are not contradictory. Thus, in principle, for two different outcomes 
of $A$ the conditional measurement can give with probability 1 the same outcome of $B.$  Such perfect conditional 
correlations cannot be symmetric, since symmetry contradict to the conclusion of section \ref{ABB}.

Consider now the case of the dichotomous observable $A,$ i.e., $E_A(+) + E_A(-) =I.$ Then, (\ref{Lx8n}) implies, that  
$ (E_B(\beta) E_A(+) + E_B(\beta) E_A(-)) \vert \psi \rangle = \vert \psi \rangle,$  i.e., 
\begin{equation}
\label{Lx8f}
E_B(\beta)  \vert \psi \rangle = \vert \psi \rangle,
\end{equation}
i.e., $\vert \psi \rangle \in  {\cal H}_B(\beta).$  This is the necessary condition of $(+, \beta)$ and $(-, \beta)$ perfect correlations  for a dichotomous observable $A.$ This condition jointly with (\ref{Lx8n}) imply that 
$
E_B(\beta) E_A(\gamma) \vert \psi \rangle=E_A(\gamma) E_B(\beta)  \vert \psi \rangle, \; \gamma=  \pm 
$
i.e.,
\begin{equation}
\label{tt77}
[E_B(\beta), E_A(\gamma)] \vert \psi \rangle =0.
\end{equation}
Thus, the corresponding projectors commute on this state.  

Consider now the case of projectors commuting on some $\beta$-eigenstate of $B,$ i.e., (\ref{Lx8f}) and (\ref{tt77}) hold. In this case,
$E_A(\gamma) E_B(\beta)  \vert \psi \rangle =  E_B(\beta)  E_A(\gamma) \vert \psi \rangle= E_A(\gamma) \vert \psi \rangle$ and, 
hence, (\ref{Lx8n}) hold.   

In particular, for commuting operators, $[\hat A, \hat B]=0,$ and dichotomous $A,$ a state is  $(\pm, \beta)$ EPR entangled iff      
it is the $\beta$-eigenstate of $B.$ Of course, this state should satisfy the condition $E_A(\gamma) \vert \psi \rangle \not=0,
\gamma= \pm.$ We remark that if both observables are dichotomous to such entanglement cannot be symmetric. Thus condition of entanglement symmetry excludes the `pathological cases"'.

Now turn to section \ref{H4}. Let $\{e_1,e_2,e_3, e_4\}$ be some orthonormal basis in  ${\cal H}.$ 
Let ${\cal H}_A(\pm)$  and ${\cal H}_B(\pm)$ be generated by $\{e_1,e_2\}, \{e_3, e_4\}$ and $\{e_1,e_3\}, \{e_2, e_4\},$ 
respectively. A state $\vert \psi \rangle= d_1 e_1 + d_3 e_3$ is the $B=+$ eigenstate and $E_A( \pm) \vert \psi \rangle\not=0.$ 
So, this state is both $(A=+, B=+)$ and $(A=-, B=+)$ perfectly conditionally ($A$ then $B)$ correlated. We can check this directly,
$$
P(B=+| A=\pm,  \vert \psi \rangle) = \Vert E_B(+) E_A(\pm) \vert \psi \rangle\Vert^2/ \Vert E_A(\pm) \vert \psi \rangle\Vert^2
$$
$$
=\Vert E_A(\pm) \vert \psi \rangle\Vert^2/ \Vert E_A(\pm) \vert \psi \rangle\Vert^2= 1.
$$

However, as we know, such entanglement cannot be symmetric. We again check this directly:
$$
P(A=\pm | B=+,   \vert \psi \rangle) = \Vert E_A(\pm) E_B(+) \vert \psi \rangle\Vert^2/ \Vert E_B(+)  \vert \psi \rangle\Vert^2 
$$
$$
=\Vert E_A(\pm) E_B(+) \vert \psi \rangle\Vert^2\not=1.
$$

\section{EPR entangled states and non-commutativity }
\label{SSS}

Consider now two pairs of dichotomous observables $A_i,B_j = \pm 1, i, j =1,2,$ such that observables in each pair 
$(A_i,B_j)$ are compatible, i.e., 
$[\hat A_i, \hat B_j]=0, i,j=1,2.$  We search the states inducing the combination of 
the perfect correlations $(A_i=\pm, B_i=\mp), i=1,2.$ They are given by the equalities:
\begin{equation}
\label{Lx8f1}
E_{B_1}(+)  \vert \psi \rangle = E_{A_1}(-) \vert \psi \rangle, 
\; E_{B_2}(+)  \vert \psi \rangle = E_{A_2}(-) \vert \psi \rangle
\end{equation}
Hence, 
\begin{equation}
\label{Lf1}
E_{B_2}(+) E_{B_1}(+)  \vert \psi \rangle =  E_{B_2}(+)  E_{A_1}(-)  \vert \psi \rangle = 
E_{A_1}(-) E_{B_2}(+)  \vert \psi \rangle = E_{A_1}(-) E_{A_2}(-) \vert \psi \rangle,
\end{equation}
in the same way
\begin{equation}
\label{Lf2}
E_{B_1}(+) E_{B_2}(+)  \vert \psi \rangle =  E_{A_2}(-) E_{A_1}(-) \vert \psi \rangle,
\end{equation}
thus, 
\begin{equation}
\label{Lf3}
[E_{B_2}(+), E_{B_1}(+)]  \vert \psi \rangle =   [E_{A_1}(-), E_{A_2}(-)] \vert \psi \rangle,
\end{equation}
We also have
\begin{equation}
\label{Lx8f1bb}
E_{B_1}(-)  \vert \psi \rangle = E_{A_1}(+) \vert \psi \rangle, \; E_{B_2}(-)  \vert \psi \rangle = E_{A_2}(+) \vert \psi \rangle,
\end{equation}
thus 
\begin{equation}
\label{Lf3nn}
[E_{B_2}(-), E_{B_1}(-)]  \vert \psi \rangle =   [E_{A_1}(+), E_{A_2}(+)] \vert \psi \rangle,
\end{equation}
Finally, (\ref{Lx8f1}) implies the equalities for combinations $(+,-)$:
\begin{equation}
\label{Lf5}
E_{B_2}(-)  E_{B_1}(+)  \vert \psi \rangle = E_{A_1}(-) E_{A_2}(+) \vert \psi \rangle , 
\end{equation}
\begin{equation}
\label{Lf5nn}
E_{B_1}(+)  E_{B_2}(-)  \vert \psi \rangle = E_{A_2}(+) E_{A_1}(-) \vert \psi \rangle , 
\end{equation}
In turn they imply that 
\begin{equation}
\label{Lf6}
[E_{B_2}(-), E_{B_1}(+)]  \vert \psi \rangle =   [E_{A_1}(-), E_{A_2}(+)] \vert \psi \rangle,
\end{equation}
In the same way we obtain that  
\begin{equation}
\label{Lf7}
[E_{B_2}(+), E_{B_1}(-)]  \vert \psi \rangle =   [E_{A_1}(+), E_{A_2}(-)] \vert \psi \rangle,
\end{equation}
Finally we get
\begin{equation}
\label{Lf8}
- [\hat B_1,\hat B_2]  \vert \psi \rangle =   [\hat A_1,\hat A_2] \vert \psi \rangle.
\end{equation}
In particular, for correlation, we get 
\begin{equation}
\label{Lf8a}
- \langle \psi| [\hat B_1,\hat B_2]  \vert \psi \rangle =  \langle \psi| [\hat A_1,\hat A_2] \vert \psi \rangle.
\end{equation}
This is the good place to make the following remark. In the tensor product situation (section \ref{TPS}), the operator $\hat b$ need not be selected as $\hat b= - \hat a,$ i.e., $\hat B$ need not be equal to $- I \otimes \hat a.$ This is an arbitrary operator. 

Finally, we notice that 
 \begin{equation}
\label{Lf8mmm}
- [\hat A_1,\hat A_2] [\hat B_1,\hat B_2]  \vert \psi \rangle =   [\hat A_1,\hat A_2]^2 \vert \psi \rangle.
\end{equation}
 and, hence, 
\begin{equation}
\label{Lf11}
-\langle \psi| \; [\hat A_1,\hat A_2] [\hat B_1,\hat B_2] \; \vert \psi \rangle =   ||\; [\hat A_1,\hat A_2] \vert \psi \rangle \; ||^2 
\end{equation}

\section{Existence of EPR entangled states for families of pairs of commuting operators}

We consider the scheme of sections \ref{SSS}. The natural question arises on the existence of the solutions of the system of 
linear equations (\ref{Lx8f1}). More generally one can consider a family of pairs of operators $(\hat A_u, \hat B_u),$ where $u$ is 
some parameter. Can one find a quantum state which is EPR entangled for all these pairs?   

For simplicity, it is assumed that, for any $u,$  $[\hat A_u, \hat B_u] =0,$ but it may be that, for some pairs $u,v,$
$[\hat A_u, \hat A_v] \not =0$ or (and)  $[\hat B_u, \hat B_v] \not =0.$ 

Let us consider the tensor  product case ${\cal H}= H \otimes H,$ where $\rm{dim} \; H=2,$ and two operators in $H, \hat a, \hat b,$  with the eigenbases $(f_+, f_-)$ and $(g_+, g_-),$ where $\hat a f_{\pm}=     f_{\pm}, \hat b g_{\pm}=     g_{\pm}.$ Set n$\hat A= \hat a \otimes I, \hat B= I \otimes b.$

Then the EPR $(\pm, \mp)$ entangled state has the form
\begin{equation}
\label{Ltax}
\vert \psi \rangle =  c_{+-} \vert f_+ g_-\rangle + c_{-+} \vert f_- g_+\rangle,\; c_{-+}, c_{-+} \not=0.
\end{equation}
This is the good place to make the following foundational remark. Set $c_{+-}= -c_{-+} (=1/sqrt{2}),$ so
\begin{equation}
\label{Ltayt}
\vert \psi \rangle =  (\vert f_+ g_-\rangle - \vert f_- g_+\rangle)/\sqrt{2}.   
\end{equation}
This state looks as the singlet state. However, the situation is really delicate. In fact, this is not the conventionally considered singlet state having the form:  
\begin{equation}
\label{Ltayz}
\vert \psi \rangle =  (\vert f_+ f_-\rangle - \vert f_- f_+\rangle)/\sqrt{2},   
\end{equation}
i.e., it corresponds to the EPR $(\pm,\mp)$ entanglement for $\hat a = \hat b,$ i.e., $\hat A=  \hat a \otimes I, 
\hat B= I \otimes \hat a.$ Nevertheless, state (\ref{Ltayt}) has the same basic property as the state (\ref{Ltayz}), 
namely, it preserves its form under tensor product of unitary transformation in $H$ with itself.   

Take any unitary transformation $\hat u$  in $H$ and consider operators of the form:
\begin{equation}
\label{A77}
 \hat a_u =  \hat u \hat a \hat u^\star, \hat b_u =  
\hat u \hat b \hat u^\star,
\end{equation}
which are diagonal w.r.t. to the bases  $(f_+', f_-')$ and $(g_+', g_-'),$  obtained with the unitary transformation $\hat u$ from the bases $(f_+, f_-)$ and $(g_+, g_-).$ Then the state $\vert \psi\rangle$ preserves its form 
if and only if $c_{+-}= - c_{-+},$ i.e., 
\begin{equation}
\label{A77a}
\vert \psi \rangle =  (\vert f_+ g_-\rangle - \vert f_- g_+\rangle)/ \sqrt{2}= (\vert f_+' g_-'\rangle - 
\vert f_-' g_+'\rangle)/ \sqrt{2}. 
\end{equation}
Hence, such state is EPR $(\pm, \mp)$ entangled both for the pair $\hat A, \hat B$ and $\hat A_u=\hat a_u\otimes I , 
\hat B_u= I \otimes \hat b_u$ (so for all pairs $(\hat A_u, \hat B_u)).$ 

\section{Brief foundational discussion}
\label{BFD}

This paper is devoted to an alternative mathematical approach to the notion of entanglement, so not directly towards quantum foundations.
However, it may be useful to complete it by a brief foundational discussion. In this section we continue to consider the tensor product state space and operators of the form $\hat A= \hat a \otimes I, \hat B=I \otimes \hat b$ and the EPR entanglement for $A=- B,$ i.e.,
$(\pm, \mp).$ These operators represent ``local observables'' $A$ and $B$ on the subsystems $S_1$ and $S_2$ of a 
compound system $S=(S_1,S_2).$  

\subsection{Elements of reality vs. perfect conditional correlations}

In the PCC approach, for any pair of operators $(\hat A, \hat B),$ there exists  $A= - B$ entangled state $|\psi \rangle \equiv |\psi \rangle_{AB}.$ Thus by getting the outcome $A= \alpha$ for  $A$-measurement on the system $S_1,$ one can with probability 1 predict the outcome $B= - \alpha$ for  $B$-measurement on the system $S_2$ without disturbing it in any way. Thus, according to the EPR notion of an element of reality \cite{EPR} $B= - \alpha$ has to be treated as the element of reality. 

However, in our formalism the case $\hat a= \hat b,$ 
i.e., $\hat A= \hat a \otimes I, \hat B=I \otimes \hat a,$ is only the special case of the general formalism with 
arbitrary operators $\hat a, \hat b.$ By ignoring the finite dimensional treatment of the problem in the present paper, we call 
observables $a$ and $b$ ``position'' and ``momentum'' of  $S_1$ and $S_2,$ respectively. 
But, it seems to be unnatural to speak about reality of ``momentum'' of the system $S_2$  by determining ``position'' of 
the system $S_1.$  It is more natural to treat this situation in the spirit of Schr\"odinger \cite{SCHE, SCHE1}
(as well as Bohr \cite{BR}), or in the rigorous mathematical terms, in the conditional probability framework.

Hence, embedding of the EPR argument in more general framework (PCC for arbitrary pairs $(A,B)$ 
of observables) moves us from the  EPR coupling of the PCC with elements of reality. 
The EPR entangled states are states describing the possibility of predictions 
with probability 1 (cf. Plotnitsky \cite{PL,PL1}). 
        
\subsection{Bell vs. EPR: joint vs. conditional measurements}  

Typically the Bell argument \cite{Bell0} is considered as just the new mathematical restructuring of the original EPR argument \cite{EPR}.
However, a few authors questioned this viewpoint (e.g., \cite{Muynck,TF}). Mathematical formalization of the notion of entanglement on the 
basis of conditional probability illuminates the difference between Bell and EPR arguments, as the difference between joint and conditional measurements.  

To speak about joint measurements in the rigorous terms, one has to appeal to von Neumann mathematical formalization \cite{VN} of the notion of the joint measurement of observables $A$ and $B.$ It must be defined an observable $K$ such that 
$A=f(K), B=g(K)$ and, for corresponding operators $\hat A=f(\hat K), \hat B=g(\hat K).$ Such observable $K$ cannot be local, it is nonlocal by its meaning. However, this nonlocality has the simple classical structure based on the introduction of the time window 
and identification of the clicks inside the selected time window. 

In contrast, conditional measurements are local per their definition.
       	
\section{Concluding remarks}
 
We demonstrated once again  that the interpretation of quantum probability as the conditional one sets the natural probabilistic meaning to the basic quantum constructions (cf. \cite{Koopman}-\cite{BL1}). In this paper, quantum conditioning was used to embed the EPR correlations into the scheme of quantum conditioning which is mathematically described by the projection postulate (in the L\"uders form). This gives the possibility to treat these correlations as entanglement of knowledge which can be extracted via conditional measurements. In this way   the notion of entanglement can be decoupled from the compound systems (and mathematically from the tensor product structure)  and, hence, from and quantum nonlocality (``spooky action at a distance''). 

Our approach to entanglement matches the original Schr\"odinger viewpoint on it, as ``entanglement of predictions'',  ``entanglement of our knowledge'', see \cite{SCHE1}:

\medskip

{\it ``... between these two systems an entanglement can arise, which ... can be compactly shown in the two equations: $q = Q$ and $p = -P.$ That means: I know, if a measurement of $q$ on the system yields a certain value, that a $Q$-measurement performed immediately 
there-after on the second system will give the same value, and vice versa; and I know, if a $p$-measurement on the first system yields a certain value, that a $P$-measurement performed immediately thereafter will give the opposite value, and vice verse.'' } 

\section*{Appendix A: Entanglement of operator algebras}

Articles \cite{Z1,Z2} describe the framework which can be refereed as operator algebras entanglement; we cite again \cite{Z2}:

``Our definitions will be observable-based and will mostly involve algebraic objects. Let us consider a quantum
system with finite-dimensional state-space $H,$ a subspace $C \subset  H,$ and a collection $\{{\cal A}_i\}_{i=1}^n$
of subalgebras of $\rm{End}(C)$ satisfying the following three axioms:
\begin{itemize}
\item i) Local accessibility: Each ${\cal A}_i$ corresponds to a set of
controllable observables.
\item ii) Subsystem independence: $[{\cal A}_i, {\cal A}_j ] = 0, i \not= j.$
\item iii) Completeness: $\cup_{i=1}^n{\cal A}_i \cong \otimes_{i=1}^n {\cal A}_i \cong \rm{End}(C).''$
\end{itemize}
This axiomatization  of the observational entanglement led to the following basic result \cite{Z1,Z2}; here abbreviation
TPS is used for a {\it tensor product structure.}  

\medskip

{\bf Proposition 1.}  A set of subalgebras ${\cal A}_i$ satisfying 
Axioms i)–iii) induces a TPS  $C = \otimes_{i=1}^n  H_i.$ We call such a
multi-partition an induced TPS.

\medskip

This proposition connects the algebraic definition with the  standard state based definition and consideration of subspace
$C$  extends essentially the domain of applicability.  Reconstruction of TPS $C = \otimes_{i=1}^n  H_i$ is very important for applications to  quantum information \cite{Za1})  and sets  coupling with system decompositions or superselection symmetries (see \cite{Za2,Za3}). 

At the same time reduction of iii) to TPS on $C$ highlights again the role of the tensor product structure. (The crucial axiom behind 
Proposition 1 is iii).) 

The conditional probabilistic approach to observational entanglement presented in our paper is a good foundational complement to 
algebraic studies in articles \cite{Z1,Z2}. We also point out to the following algebraic type differences between two approaches.
The main difference is that we proceed without axiom iii). Generally we neither appeal to axiom ii), but the most interesting results
are obtained in the commuting case. Proceeding without appealing to axiom iii) gives us the possibility to disconnect entanglement completely from TPSs, but reserve the possibility of such connection in special cases. 

On the other hand, we consider the special form of observational entanglement. This form matches the framework of the EPR-paper \cite{EPR} and mathematically formalizes  perfect correlations between the outcomes of observables -- 
via conditioning with probability $p=1.$ Such unit probability correlations are important from the foundational viewpoint for 
rigorous treatment of the EPR elements of reality \cite{EPR}.  

\section*{Appendix B: Multi-observables EPR entanglement}

Consider now three discrete quantum observables $A,B, C$ represented by Hermitian operators $\hat A, \hat B, \hat C$ and generating the projection type state update. We are interested in conditional probability
to get the outcome $C= \gamma$ if  the preceding sequential measurements, first $A$ and then $B,$  the outcomes $A=\alpha$ and 
$B=\beta$ were obtained: 
\begin{equation}
\label{A1}
P(C= \gamma | A= \alpha, B=  \beta, \psi)= \frac{\Vert  E_C(\gamma) E_B(\beta) E_A(\alpha) \vert \psi \rangle\Vert^2}{\Vert  E_B(\beta)   E_A(\alpha) \vert \psi \rangle \Vert^2}.
\end{equation}
The outcome $C= \gamma$ is perfectly correlated with the (previous) outcomes $A=\alpha$ and  $B=\beta$ if this conditional probability equals to 1, i.e., 
\begin{equation}
\label{A2}
\Vert  E_C(\gamma) E_B(\beta) E_A(\alpha) \vert \psi \rangle\Vert^2 = \Vert  E_B(\beta)   E_A(\alpha) \vert \psi \rangle \Vert^2.
\end{equation}
Hence, 
\begin{equation}
\label{A3}
E_C(\gamma) E_B(\beta) E_A(\alpha) \vert \psi \rangle =   E_B(\beta)   E_A(\alpha) \vert \psi \rangle .
\end{equation}
Of course, the condition of non-degeneration must hold:
\begin{equation}
\label{A4}
E_B(\beta)  E_A(\alpha) \vert \psi \rangle \not=0.
\end{equation}
Hence,  (\ref{A3}), (\ref{A4}) are necessary and sufficient conditions of the perfect correlation $(A= \alpha, B=\beta, C=\gamma).$

Consider now the case of compatible observables $A$ and $B.$  Ee remark that in this case $(A= \alpha, B=\beta, C=\gamma)$ and $(B=\beta, A= \alpha,  C=\gamma).$ perfect correlations are  equivalent.

Now let all observables be compatible;  we find the conditions of the joint perfect correlations,  
$(A= \alpha, B=\beta, C=\gamma), (B=\beta, C=\gamma, A= \alpha), (C=\gamma, A= \alpha,B=\beta),$
\begin{equation}
\label{A5}
E_C(\gamma) E_B(\beta) E_A(\alpha) \vert \psi \rangle =   E_B(\beta)   E_A(\alpha) \vert \psi \rangle .
\end{equation}
\begin{equation}
\label{A6}
E_A(\alpha) E_C(\gamma) E_B(\beta)  \vert \psi \rangle =   E_C(\gamma) E_B(\beta)  \vert \psi \rangle .
\end{equation}
\begin{equation}
\label{A7}
E_B(\beta)   E_A(\alpha) E_C(\gamma)\vert \psi \rangle =   E_A(\alpha) E_C(\gamma)\vert \psi \rangle.
\end{equation}
Thus, 
\begin{equation}
\label{A8}
 E_B(\beta)   E_A(\alpha) \vert \psi \rangle = E_C(\gamma) E_B(\beta)  \vert \psi \rangle = E_A(\alpha) E_C(\gamma)\vert \psi \rangle,
\end{equation}

Consider observables on a compound system $S=(S_1, S_2, S_3)$ with the state space 
${\cal H}= {\cal H}_1\otimes {\cal H}_2 \otimes {\cal H}_3;$ let $a, b,c$ be local observables
with operators $\hat a, \hat b, \hat c$ and let $A, B, C$ be the corresponding observables 
on $S$ with operators    $\hat A=\hat  a \otimes I \otimes I, \hat B= I  \otimes \hat b \otimes I, 
\hat C=  I \otimes I \otimes  \hat c.$ Suppose that all observables are dichotomous with values $\pm 1.$ 
The standard basis in ${\cal H}$ is given by $(|\alpha \beta \gamma\rangle).$ 
The general state is represented as
$$
|\psi \rangle = \sum_{\alpha, \beta, \gamma = \pm} c_{\alpha \beta \gamma}  |\alpha \beta \gamma\rangle.
$$
We now describe the PCC-states for $(A=\pm, B=\pm, C = \pm),$``EPR entangled'' for the set triples $(+++), (---).$
The system of equalities (\ref{A8}) implies, for $\alpha, \beta, \gamma =\pm,$
$$
  c_{+ + +}  |+ + +\rangle + c_{+ + -}  |+ + -\rangle =  c_{+ + +}  |+ + +\rangle + c_{- + +}  |- + +\rangle= 
$$
$$
c_{+ + +}  |+ + +\rangle + c_{+ - +}  |+ - +\rangle
$$
Hence $c_{+ + -}= c_{- + +} = c_{+ - +} =0.$ In the same way, by selecting $\alpha, \beta, \gamma = -,$
we get $c_{+ - -}= c_{- - +} = c_{- +-} =0.$ Hence
\begin{equation}
\label{A9}
c_{+ + +}  |+ + +\rangle + c_{- - -}  |- - -\rangle, \; c_{+ + +}, c_{---} \not=0.
\end{equation}
and with normalization by 1.

\end{document}